# Design and Microfabrication Considerations for Reliable Flexible Intracortical Implants

H S Sohal[1], K Vassilevski[3], A Jackson[2], S N Baker[2], and A O'Neill[3]

[1]Media Lab, McGovern Institute, Microsystems Technology Laboratories, MIT, Cambridge, MA, 02141, USA

[2]Institute of Neuroscience, Newcastle University, Newcastle Upon Tyne, NE2 4HH, UK

[3] School of Electrical and Electronic Engineering, Newcastle University, Newcastle Upon Tyne, NE1 7RU, UK

E-mail: hssohal@mit.edu

**Abstract**

Current microelectrodes designed to record chronic neural activity suffer from recording instabilities due to the modulus mismatch between the electrode materials and the brain. We sought to address this by microfabricating a novel flexible neural probe. Our probe was fabricated from parylene-C with a WTi metal, using contact photolithography and reactive ion etching, with three design features to address this modulus mismatch: a sinusoidal shaft, a rounded tip and a polyimide anchoring ball. The anchor restricts movement of the electrode recording sites and the shaft accommodates the brain motion. We successfully patterned thick metal and parylene-C layers, with a reliable device release process leading to high functional yield. This novel reliably microfabricated probe can record stable neural activity for up to two years without delamination, surpassing the current state-of-the-art intracortical probes. This challenges recent concerns that have been raised over the long-term reliability of chronic implants when Parylene-C is used as an insulator, for both research and human applications The microfabrication and design considerations provided in this manuscript may aid in the future development of flexible devices for biomedical applications.

# 1. Introduction

Chronically implanted intracortical microelectrodes have been used in invasive brain machine interfaces (BMI) for controlling computers and assistive devices to restore function after paralysis[1–3]. However, current devices suffer from recording instabilities that lead to loss of cells over time[4]. One reason for this is the Young's modulus mismatch between electrode materials and brain tissue, leading to micromotion induced trauma [5,6] and a persistent immune response that compromises neural function [7].

One way to mitigate the Young's modulus (E) mismatch is to produce flexible interfaces for long-term implants. Current flexible electrodes are centered on the use of polymers such as Parylene-C (E = 2.86 GPa), polyimide (E ~ 2.3 GPa), SU-8 (E = 4.02 GPa) and smart materials with a suitable metal for recording, although chronic recording data is lacking. Most of the current flexible devices seem to develop issues, especially for interfacing > 1 year. For example, polyimide has been used extensively for flexible electrodes [8,9], however it suffers from pin-hole development, which will lead to subsequent *in-vivo* device delamination through fluid infiltration and is also approximately 3 times stiffer (in the case of PI-2611) than Parylene-C based electrodes [10]. Additionally, performing leak impedance measurements after submerging PI-2611 in 40°C saline found sharp declines in impedances after 2-3 days, alluding to insulation failure on probes, which would translate to poor *in-vivo* performance. However, enhanced performance can be obtained, if specific curing measures are used [11]. Smart materials [12–14] that change E when being inserted due to water uptake or a change in temperature (e.g. 5 GPa 12 MPa would be an ideal solution) suffer from poor insulating properties and have only been used as insertion aids for flexible structures or lack major characterization beyond a proof-of–principle demonstration [12]. Further, chronic implantation after 8 weeks showed a similar distribution of defect nuclei compared to a microwire electrode, therefore conferring no advantage to prolong chronic recording through modulus matching with these materials [13,15]. Conductive polymers [16] and carbon fibers have also been used, however these lack *in-vivo* characterization for indwelling periods > 1 year [17] and seem to confer no advantage over conventional electrodes in terms of gliosis [18]. Finally, carbon fiber probes, which can be produced at small diameters with E = 241 GPa [19], can be more stiff than inflexible silicon probes, although good chronic recording has still been obtained with such probes [20]

Here, we describe a microfabrication process flow that led to high functional yields (>95%) in post-processing. Using our devices, we were able to record neural activity for > 2 year periods, demonstrating the long-term reliability of our microfabricated devices [21]. The probe comprised a parylene-C-metal-parylene-C sandwich with design features to minimize the effects of micromotion. We report design and microfabrication considerations that allowed reliable devices to survive the challenging environment of chronic neural recording [22]. Recording and gliosis responses to our devices have been reported elsewhere [21]. A combination of overall design measures, optimized microfabrication and reduced gliosis may have contributed to the enhanced longevity of the probe.

## 2. Design

Compared with conventional rigid and sharp electrodes, the design had several features to minimize micromotion-induced trauma (Figure 1A). The probe was made out of flexible materials and incorporated a rounded tip with three protruding recording sites, which were anchored within the brain tissue with a polyimide ball. Further, the design incorporated a sinusoidal shaft, instead of a straight shaft. This shaft accommodates movement of the brain, mechanically decoupling the recording end from the surface fixation point and thus allowing the electrode sites and ball anchor to move freely with the tissue. The recording end of the probe was a Parylene-C disc of 100 μm diameter, with three protruding, exposed recording sites (96 μm$^2$) to aid single-unit isolation. Three metal tracks (5 μm width and gap) ran within an electrode shaft that was 20 μm deep, 35 μm wide and had a sinusoidal profile with cycles of 100 μm amplitude and 500 μm period. The probe was 3 mm in length and incorporated an integrated Parylene-C based ribbon cable (3 cm long and 3 mm wide) leading to a standard connector (micro ps1/ps2 series, Omnetics Connector Corporation, USA) (Figure 1B,C). To demonstrate the utility of a sinusoidal shaft a finite element model (FEM) was constructed (Figure 2). GDS files from mask designs were imported into COMSOL Multiphysics®, for both straight and sinusoidal shafts with identical dimensions (3 mm length, 20 μm thick, 35 μm wide) to form a 3D model. The "Solid Mechanics" physics was used and the top of the electrode shaft was rigid ("fixed constraint"). Parylene-C was used in this simulation with properties of E = 2.86 GPa, density = 1.29 kg/m$^3$ and Poisson's ratio = 0.4. The shaft itself was free to move. We compared the flexibility of both designs to accommodate displacements in 3 dimensions. An arbitrary displacement (100 μm) was applied in the orthogonal X, Y, and Z directions of the rounded recording end for both the straight and sinusoidal probe design. Figure 2 shows that a sinusoidal shaft, rather than a straight shaft, will deform more to counter movement of the probe primarily in the Y direction (Average displacement of the probes (±Standard Error of the Mean (S.E.M) 85 ± 9 vs. 53 ± 9 μm$^3$, t-test: t = 5.5, P < 0.001) as calculated through COMSOL Multiphysics® using the "domain probe" feature, which allows the measurement of displacement values for the duration of the simulation. Comparable average displacement values (53.3 μm) were obtained for X and Z directions. The Y-direction has also been shown to be the primary axis of motion (cephalocaudal) for the brain in both rodents and human MR studies [23–25].

The simulation allowed the assessment of the Von Mises stress on both shank types. The von Mises stress is used to predict yielding of materials under any loading condition from results of simple uniaxial tensile tests. The stress induced by the movement was significantly reduced for a sinusoidal (30.7 MPa) rather than a straight shaft (92.5 MPa), showing that it is also better at minimizing stresses on the shank and at accommodating random brain motion.

## 3. Fabrication

### *3.1 Overall Process Flow*

Fabrication began with deposition of 1 μm thick Al on a silicon carrier wafer by e-beam evaporation. The three inch Si wafer had a smooth polished surface to aid device release following processing (Figure 3A). Then the first layer of Parylene-C was deposited with a thickness of 10 μm using a commercial service (Paratech Coating, Ltd, UK) (Figure 3B). Next, a 1 μm thick WTi (80/20 wt%) layer was deposited by magnetron sputtering and patterned using contact lithography and reactive ion etching (RIE) in $SF_6$ containing plasma (Figure 3C). A slight over-etch was employed to roughen the surface of the first Parylene-C layer to aid the Parylene-to-Parylene adhesion. Next, the second Parylene-C layer was deposited (Figure 3D). Subsequently Parylene-C layers were patterned using contact lithography and RIE in oxygen plasma with a Ti mask (Figure 3E). The RIE process was optimized to leave the exposed WTi unaltered and produce a near-vertical sidewall profile in Parylene-C. Finally, the devices were released by Al etching with diluted tetramethylammonium hydroxide (TMAH) (Figure 3F).

*3.2 Metallisation*

Tungsten-titanium alloy (WTi) was used as the metal layer. Tungsten and has been used extensively for the past 50 years to record from the brain [26] but is relatively stiff compared to other metals, with a modulus of elasticity of 400 GPa. Therefore tungsten was sintered with titanium using a combined metal sputtering target for increased flexibility (Young's modulus: 110 GPa).

To form conductive tracks in the electrodes, a 1 μm thick WTi layer was deposited by magnetron sputtering in a Kurt Lesker PVD 75 vacuum deposition system using a sintered target with WTi (80/20 wt%) from Pi-Kem Ltd. An unconventional 1 μm thick WTi metal was used for the tip based recording sites, where the thickness of the metal defines the recording site area. The electrode recording sites are 4 exposed faces of a cuboid. Overall, a 1 um thick layer will yield a recording area of around 96 $μm^2$. Thinning the metal down to 500 nm will give a site area of 80 $μm^2$. A thicker metal would also be more resistant to potential chronic degradation of the metal due to electrochemistry, a further design measure used to enhance probe longevity [27].

RIE in $SF_6$ plasma (Figure 4) was chosen instead of a wet etch as it provided a means to surface roughen the first Parylene-C layer (improving overall Parylene-C adhesion), by employing a simple "over-etch." Also, the RIE process has the advantage that it reduces the risk of Parylene-C swelling during wet etching [28]. A novel wet etch was also performed with a mixture of ammonia hydroxide (30%) and hydrogen peroxide (30%) (1:1 ratio) and it was found that 1 μm thick WTi (80/20 wt%) was etched in 5 minutes. A 3 μm thick TI-35ES photoresist (HD Microchemicals, Germany) was used as a mask for this process. Optical microscopy (observed signs of pitting and delamination) and multi-meter measurements confirmed that the Parylene-C layer remained unaffected. This etch duration was shorter than the use of conventional hydrogen peroxide (30% ~ 40 minutes).

*3.3 Thick parylene-C etching*

The chosen dielectric was parylene-C, due to excellent mechanical and electrical properties, including a low modulus of elasticity (2.86 GPa), and elongation at break of 200%, suitable to accomodate large deflections. Moreover, parylene-C has a high long-term dielectric constant due to a very low water uptake at 50% humidity [29–31]. Importantly, parylene-C already has FDA approval for use in a biomedical device in human intracortical implants [1,2] and is used as the insulator for the wiring on cardiac pacemakers. Also parylene-C has obtained the highest biocompatibility rating; class VI, from the United States Pharmacopeia (USP)[1,32]. Additionally to published literature, the film quality was observed between polyimide and parylene-C post-deposition. PI-5878G (HD MicroSystems, USA) was chosen to be a comparison dielectric to parylene-C due to similar electrical and mechanical properties. The polyimide was spin deposited on to a three-inch Si (300 μm thickness; single sided polished) wafer coated with an aluminium sacrificial layer via pastette deposition. To achieve a similar thickness as parylene-C, the polyimide had to be deposited in two layers with a soft-baking step in between. The polyimide was then soft-baked in accordance to manufacturers guidelines, 120 °C for 30 minutes in a convection oven. Further films were then fully cured in a $N_2$ environment at 350 °C for 2 hours to drive away any remaining solvent and for complete imidization.

Dielectric failure has been attributed to the accumulation of bubbles within the layer (Rubehn and Stieglitz, 2007), leading to pitting over time. A bubble test was performed on polyimide and parylene-C coated on three-inch wafers with an aluminium sacrificial layer. Ten, random images were taken for layers on x 5 optical magnification. The number of bubbles per image was counted manually with the use of ImageJ64 software (NIH, USA). A t-test was then performed to compare the two groups. The polyimide layer had a significantly greater number (t= 6.2, P<0.001) of bubbles than a parylene-C layer. An overall ratio for the number of bubbles in polyimide compared to parylene-C was calculated at 8:1. Therefore, we also show that parylene-C was superior to polyimide as the dielectric has less potential for device failure due to pitting (Figure 5).

Relatively thick Parylene-C films (20 μm) were etched using RIE in oxygen plasma. This thickness was chosen as a balance between flexibility and improved chronic performance versus potential insulation degradation. Since a photoresist mask is readily etched by oxygen plasma, a metal masking layer was used. Ti (40 nm thick) was deposited by e-beam evaporation and patterned using conventional contact lithography. Titanium etching with diluted HF (HF: $H_2O$=1:60) patterned within 15 seconds (Figure 6B). AZ-5214e photoresist pre-etch baked at 115°C was used for this process and removed with acetone after etching. An Al mask was found to be unsuitable for this process due to poor Parylene-C adhesion (Figure 6A).

Wafers coated with Parylene-C and the subsequent Ti mask were subjected to RIE in oxygen plasma at room temperature, having a discharge power of 200 W, pressure of 50 mTorr and gas flow rate of 18 sccm. The RIE process provided an anisotropic etch of Parylene-C with an etch rate of 230 nm/min and good selectivity to the exposed WTi and Al. Resultant sidewalls of fabricated electrodes were vertical as shown by optical (Figure 6C) and scanning electron microscopy (Figure 6D). To determine the side wall angles and etch isotropy, samples were appropriately cleaved to check the sidewall profile with the use of optical microscopy. Cleaved, square samples were placed vertically, with the aid of an alignment piece, to allow focusing of the sidewall. Images were captured and the sidewall angles were measured from the optical images in the Matlab environment, as indicated by the red-dashed line in Figure 5C (2009a, MathWorks, USA). Parameters of 200 W and 50 mTorr pressure yielded vertical etch and had an angle of $88^{o}$ +/- 0.4 (n = 5 samples). Further, the vertical sidewalls were reconfirmed with scanning electron microscopy (Figure 6D). Parameters of 175 W and 100 mTorr pressure yielded less of a vertical etch and had an angle of $82^{o}$ +/- 0.6 (n = 7 samples). The Ti mask had minimal etching, with the Ti thickness still measuring 30-35 nm, after the etch.

*3.4 Sacrificial layer choice and device release*

First we used PVD deposited SiO2 as a sacrificial layer with HF (49%) etching for device release. Although device release was often successful, the release process took >24 hours, which risked damaging our completed probe [28,30]. Therefore, we used aluminum as a sacrificial layer. Previous studies have etched aluminum sacrificial layers using KOH (30%) to release devices [19]. However, KOH could potentially attack the WTi layer of our device, and effervescence during etching prevents visual confirmation of device release. We found that if the adhesion of the parylene-C to the aluminum was poor, a simple IPA soak could be used to release devices. However this sometimes resulted in reduced yield (~50%) due to ribbon cable releasing before the probe leading to breakage of tracks at the top of the electrode. If the adhesion was good between the aluminum and parylene-C, we found that etching with dilute TMAH (e.g. AZ-326 MIF, photoresist developer, HD microchemicals) could successfully release all devices within a 1.5 hour period without underlying damage to the parylene-C or WTi. This was substantially faster and more reliable than the use of a silicon dioxide sacrificial layer.

**4. Device Packaging**

*4.1 Connector Attachment*

Three methods were explored to bond the connector to our bond pad: gold wire wedge bonding, soldering and silver paint bonding. Gold wire wedge bonding was ruled out due to the flexibility of parylene-C, which caused both parylene-C and WTi to be scratched by the bonding machine. Soldering was also ruled out as the parylene-C melted and unsuitability to successfully attach to Tungsten. Silver paint bonding was found to be a highly reliable, quick and reproducible method of connector bonding. Two techniques were considered to insulate the silver paint bond: epoxy (Araldite, UK) and polyimide. Leak impedance measurements (IMP-1 meter, Bak Electronics, USA) were made at 1 kHz in saline to test the integrity of insulation with both methods. Electrodes were considered insulated if the leak impedance was > 5 MΩ.

We found that two layers of epoxy were sufficient to insulate the connector bond, and epoxy had the advantage over polyimide that it could be removed with N-Methyl-2-Pyrrolidone (NMP) allowing connectors to be re-used.

*4.2 Device attachment to insertion carrier*

In order to be implanted into the brain, the flexible electrode was attached temporarily to a sharp and rigid carrier using poly-ethylene-glycol (PEG: MW 6000) [33]. After insertion this was washed away with warm saline to liberate the electrode and allow removal of the carrier. For the first probe generation commercial fine steel electrodes were used (0.229 mm diameter, 2-3 μm tips; Microprobe inc, USA) as the carrier. However, difficulties were found securing the ribbon cables to the thin steel electrodes and therefore other methods were explored: 1) 30 G needles, filled with epoxy to prevent brain boring with the probe attached using PEG (Figure 7A), 2) half needles with a U-shaped cross-section ('stylet'), made by grinding half of a 25G needle using a Dremel sanding disk, with the probe placed in the groove and filled with PEG (Figure 7B).

The optimized insertion strategy used a carrier comprising a sharp steel electrode within a 25G syringe needle, secured in place with two-part epoxy. The probe with incorporated ribbon cable was then mildly thermoformed with a hot air gun (~120$^\circ$C) to straighten the ribbon cable. This allowed the connectors to be attached to the plastic part of the needle with PEG. Part of the ribbon cable was also secured to the insertion needle to minimise risk of breakage at the tapering point between the electrode and ribbon cable. The probe was attached with a thin PEG layer (Figure 7C). Using this method it was possible to attach multiple electrodes to a single carrier, allowing up to 4 probes to be implanted during a single brain penetration (Figure 7D). A standard stereotaxic set-up was used for this with a fast manual insertion realized with the adjustment of height on the stereotax. The stereotax was modified to incorporate a syringe manipulator (e.g. BE-8 ball joint manipulator from Narishige group, Japan). The dissolving of the PEG took ~ 1 minute and a further 1 minute wash with physiological saline was performed to ensure complete PEG dissolving, before liberation of the carrier from the brain. This preferred method decreased the risk of damaging the brain upon insertion due to a smaller insertion footprint.

## 5. Impedance characterization

Electrode impedances at 1 kHz in saline (IMP-1 meter, Bak Electronics, USA) were characterised for the probes (Figure 8A). A total of 426 recording sites were characterised and impedance values all fell within the acceptable range for single-unit recordings < 5MΩ [34]. Specifically, impedance values (±SEM) were 770 ±15 kΩ.

The microfabrication process therefore produced many viable electrodes towards the end of the optimization period with many probes falling in the acceptable impedance range for single unit recording. The majority of probes (> 83%) had impedances ≤ 1.2 MΩ, which is classed as good/ideal from data corresponding to the use of acute single shanked microwire probes (Microprobe inc, USA) to isolate single-unit activity [35,36]. A further 30 recording sites had impedances above 3 MΩ, and were not considered as viable probes for use for chronic implantation [34].

Electrode impedances were also fully characterized over a 180 day indwelling period for 15 electrode recording sites (Figure 8B) for the second generation probe. Although, there was a trend of an initial increase and decrease in impedance over the indwelling period, there was no significant difference for explanted devices when comparing the impedance values at 1 kHz in saline (recording day 0 vs. 180, t-test: $t = 1.7$, $P > 0.05$). The mean impedance values for recording day 0 and 180 were $1100 \pm 150$ (n = 15 sites) and $850 \pm 120$ kΩ (n=12 sites), respectively. This showed that the Parylene-C based devices were stable and survived damage, which could have been caused by the long indwelling period. The initial increase in impedance after day 0 was likely to be caused by the change in the impedance measurement interface (brain vs. physiological saline) and the natural acute trauma due to surgery [37]. The decrease (~35 day) and stabilization in electrode impedances were caused by the stabilization of the electrode-tissue interface [21].

## 6. Recording data

This probe has registered high fidelity neural activity for over a two-year period. Consistent waveforms were extracted with stable signal-to-noise ratio (SNR) and mean peak-to-peak amplitude over the recording period. The second probe generation could register neural activity for 140 day period with increasing SNR and peak-to-peak amplitude over the recording period (n=15 recording sites, 5 probes) [21]. These promising initial recordings highlight the stability of the implanted probes and emphasize the reliable microfabrication process that has been developed, which promotes device survival for long indwelling periods. As the probe can also be permanently attached to the carrier with two-part epoxy, a dual-purpose probe for both acute and chronic recording can be created. As a demonstration of this, we show that we can acutely record from the mouse cortex with good SNR (Figure 9) following Suner's method [31,45]. From example waveforms extracted (n=6, Figure 9B) we showed good mean peak-to-peak amplitude (200 +/- 50 uV), Noise (33 +/- 7 uV) and SNR (6.0 +/- 0.5). For the signal, the peak-to-peak amplitude (A) of the mean waveform was calculated. For the noise, the mean waveform was subtracted from all waveforms, with the standard deviation calculated from the resulting values. Noise was then calculated as 2x the average standard deviation ($\varepsilon$). SNR was then calculated as A/ $\varepsilon$. For 12 recording sites there was no difference in pre- and post-implant impedance value post-device cleaning with ethanol and deionized water (t-test: t=1, *P*>0.05; Figure 9C).

The probes are Parylene-C based, so they are easier to clean than conventional silicon probes that have been used for repeated acute recordings due to minimized adhesion of tissue and blood following acute probe explant [38]. Since the microfabrication and insertion device fabrication are relatively cheap in comparison to silicon and other probes, this raises the interesting possibility of a cost effective, re-usable probe for neural recording in acute and chronic preparations.

**7. Discussion**

The recording parameters showed that the probe was reliable during the long indwelling periods, indicating a highly reliable microfabrication process and the chronic reliability of Parylene-C. A change from $SiO_2$ to Al for the sacrificial layer allowed reliable release for the probe, as long as there was good adhesion between the Al and Parylene-C. Al has a tendency to develop a thin native oxide over time and this can interact with the silane (A-174) based adhesion promoter used for Parylene-C. Literature suggests that a greater oxide formation combined with silane will lead to enhanced adhesion between Parylene-C and Al [39]. Therefore to enhance Al adhesion to Parylene-C, a native oxide layer should be allowed to form on the Al through a simple delay in microfabrication processing, before Parylene-C deposition. This will ensure there is reliable good adhesion. Although some other groups do not use a sacrificial layer with Parylene-C or polyimide while demonstrating chronic usage up to 28 days [40,41], it was found here that device release was more reliable with its inclusion: devices occasionally failed to release without the sacrificial layer and also the peeling method would increase the risk of device damage and reduced functional yields.

Other groups have baked Parylene-C with a nitrogen backflow to increase adhesion between the two layers [41]. But a slight over-etch of the metal roughens the first Parylene-C layer surface leading to good device adhesion with no delamination to microfabricated, implanted and explanted devices, as corroborated through impedance measurements and Ramen spectroscopy. However, baking should be considered in future, especially with the incorporation of noble metals (e.g. platinum for chronic stimulation) to increase adhesion. Baking the Parylene-C based devices post-microfabrication between 200-300 $^{\circ}$C for 1-2 hours may increase longevity of the Parylene-C. For saline soaked samples at 37$^{\circ}$C over a 300 day period, stable DC resistances of samples were observed at 300$^{\circ}$C. Further, comparing untreated, 200$^{\circ}$C and 300$^{\circ}$C samples, it was found that the baseline Yield strength of Parylene-C can be enhanced with increasing temperature (77 to 107 MPa) [42].

A striking feature is that our devices remained stable for a two-year indwelling period. Although there have been reports of Parylene-C failure on intracortical probes, this is mainly due to deposition onto silicon. For the Utah array, thin (< 4 um) layers are used to insulate the array, and this will eventually break down due to adhesion of the Parylene-C to the silicon [43,44]. Material stresses due to the E mismatch between silicon and Parylene-C can also cause cracking [45]. No such problems were encountered with our probes when explanted, as similar impedance values were found pre and post- implantation (Figure 7B). Probes were able to record for two years without degradation in the acquired neural signal, emphasizing material stability [8]. A recent study comparing intracortical probes (Utah array, TDT microwires, neuronexus probes) using reactive oxygen species for accelerated longevity has also indicated delamination problems with polyimide insulated microwires and Parylene-C insulated silicon probes [46]. Literature suggests that Parylene-C failure mechanisms include water and salts diffusion [47,48] interfacial delamination [43], and pinholes [49]. The thick (although still flexible) Parylene-C layers and minimized damage during processing (e.g. enhancing adhesion between layers) may have contributed to this improved chronic performance. Interestingly, a reduction in the microglia response was observed around the probes for a 12 and 24 month period [21]. Microglia are known to produce toxic chemicals, such as reactive oxygen species, that will degrade electrode material over time [46]. A combination of overall design measures (Figure 1), material choice and reduced gliosis may have contributed to the enhanced longevity of the probe. Polyimide was found to contain many pinhole defects with a manual pastette deposition method and therefore reaffirmed the choice for Parylene-C use as an insulator.

Here, the acute recording qualities of the sinusoidal probe are demonstrated. The probe is permanently attached to the insulated sharp steel electrode and is further insulated from the effects of electromagnetic crosstalk by using a two-part epoxy attachment technique. Good SNR and signal amplitude are observed, which is comparable and in some cases better than conventional silicon probes that are used in acute animal preparations [50–52]. One advantage of a flexible probe over a silicon probe is the reduced complexities in microfabrication. This reduces cost and ease of cleaning, making it an attractive alternative to silicon probes for both acute and chronic applications.

In addition to our stable devices, reduced gliosis responses were shown, compared to microwire electrodes (the gold standard in terms of gliosis at the time of the study) at 6, 12 and 24 month time points at the electrode tip, with an increase in neural integration around our probe [21]. This suggests that the design features incorporated in our probe have reduced the effects of micromotion-induced trauma.

Although these probes were restricted to 3 mm in size and three electrode-recording sites, this can be up-scaled from a basic successful proof-of-concept design to allow for differing lengths and recording site arrangements. The design could potentially be adapted for multiple applications and differing species.

## 8. Conclusions

We have successfully microfabricated a reliable flexible probe for chronic neural recording, with one of the longest chronic indwelling periods (> 2 years), without delamination. We have optimized etching of thick parylene-C layers and WTi metal. We have also developed a reliable device release strategy incorporating an aluminum release layer and TMAH, leading to a high number of functional devices. Further, we have developed a relatively simple and inexpensive device insertion technique that has been reliably used to insert the flexible devices into rabbit brains. This microfabrication process flow can also be applied to construct other flexible electrode types.

**Figure Legends**

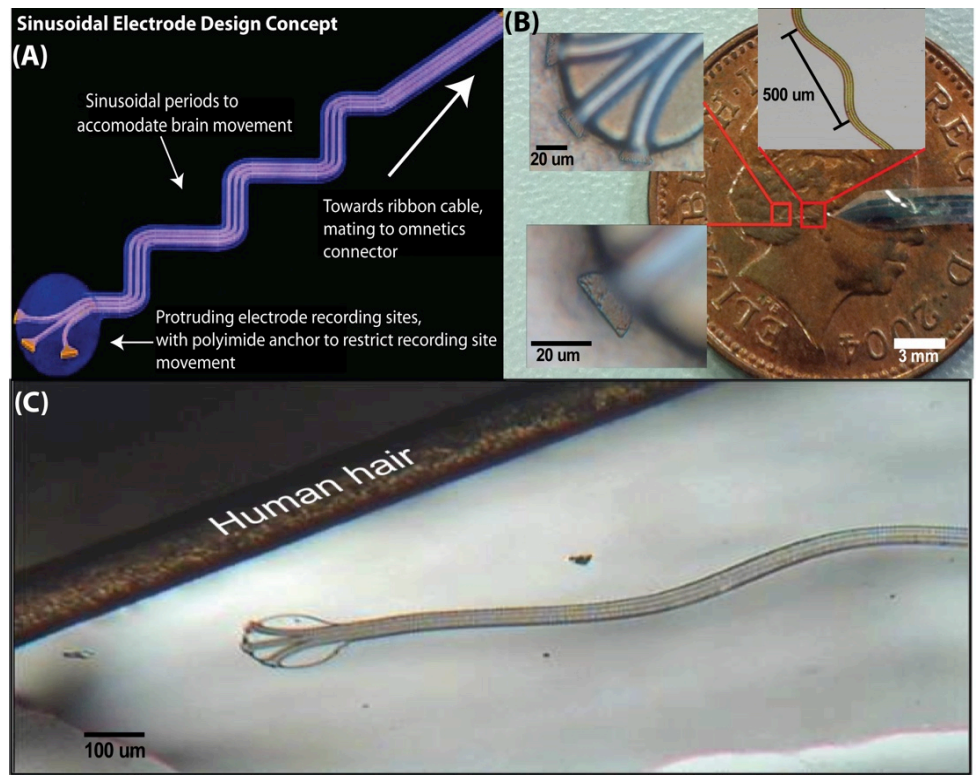

**Figure 1.** The sinusoidal probe had design measures to reduce micromotion, such as anchored recording sites and a sinusoidal shaft to accommodate brain motion (A). Microfabricated probe compared to a one-penny coin with insets showing recording sites and sinusoidal shaft, respectively (B). As a size comparison, the probe is shown against a human hair (C).

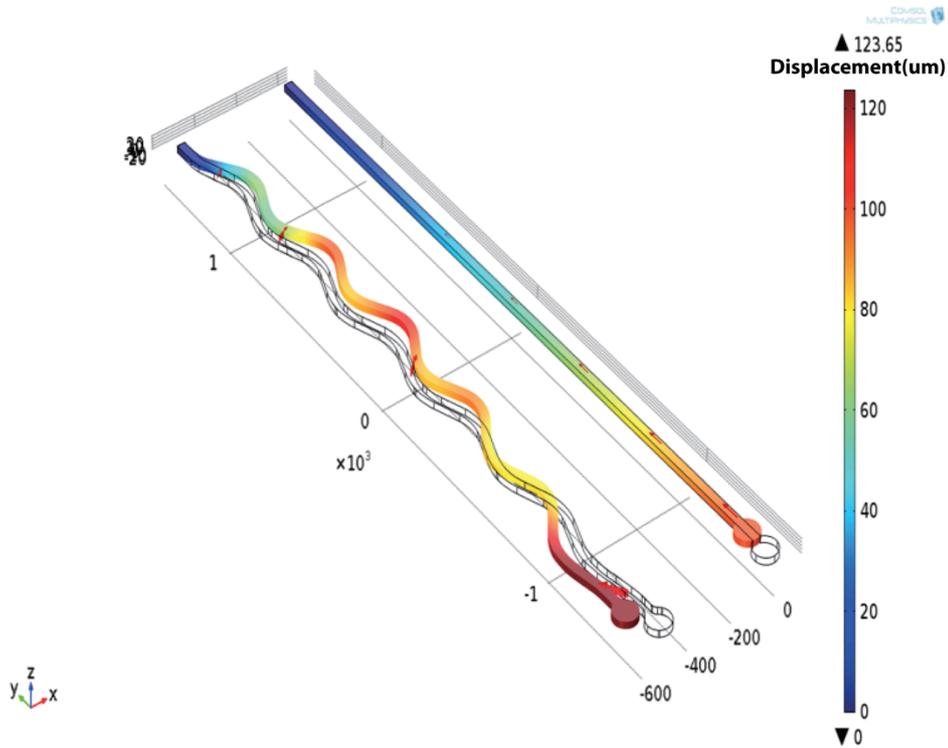

**Figure 2.** COMSOL FEM showing the utility of a sinusoidal vs. a straight shaft. The sinusoidal shaft will deform more in the Y-direction, which is thought to be the primary axis of motion in the brain.

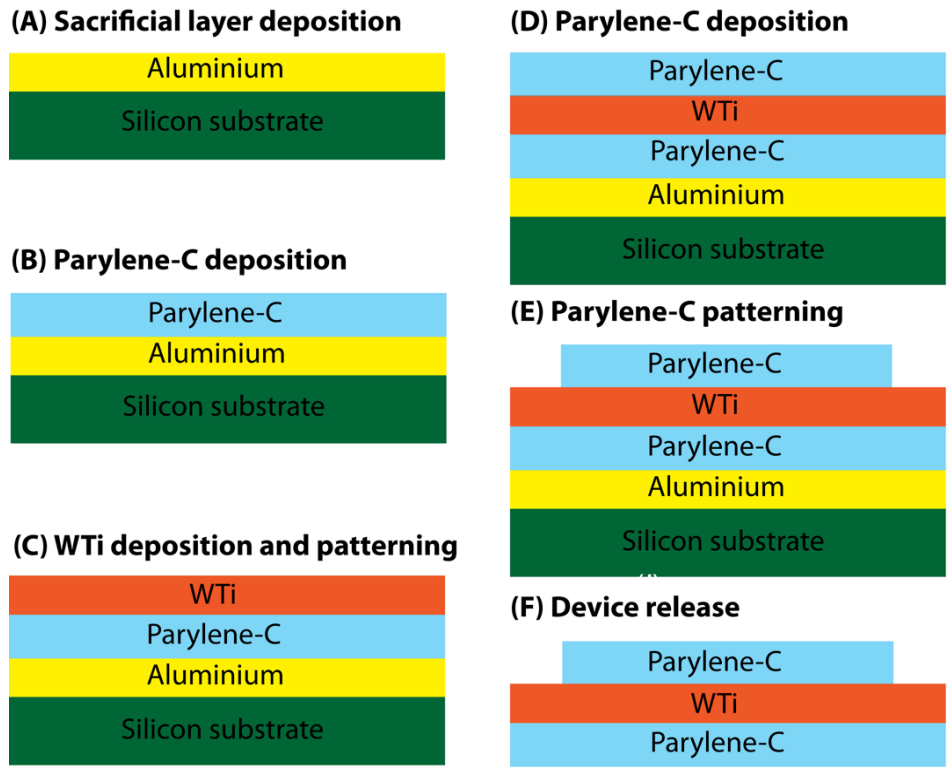

**Figure 3.** Optimised process flow for the Sinusoidal probe. Al was deposited on a three inch wafer (A) to aid device release. Sequentially, the first Parylene-C (B), WTi metal (C) and final Parylene-C layer (D,E) were deposited and patterned to form the conductive tracks and to define the overall shape of the sinusoidal probe. Finally, the device was released by etching the Al in weak TMAH (3%; F).

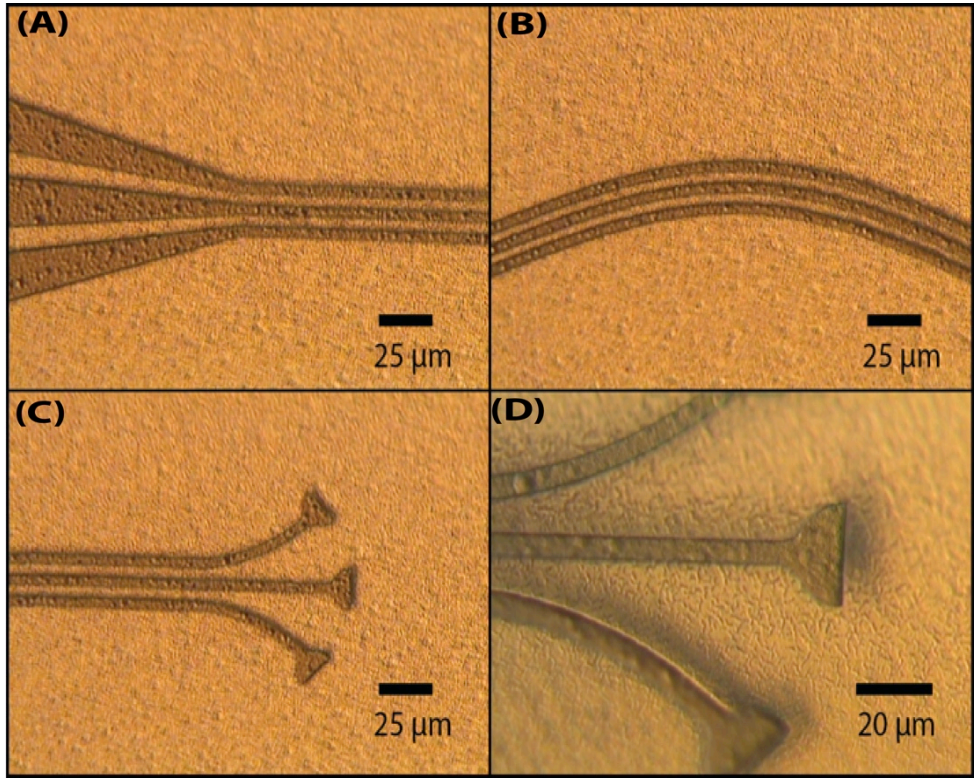

**Figure 4.** Successfully patterned WTi metal with $SF_6$ RIE. There was limited dimension loss. RIE was preferred over the wet etch due to the ability to surface roughen the first Parylene-C layer to aid Parylene-C to Parylene-C adhesion. The four optical images show various parts of the electrode design, and successful patterning of unconventional features: Ribbon cable interface with electrode tracks (A), sinusoidal electrode tracks (B), and recording sites (C,D).

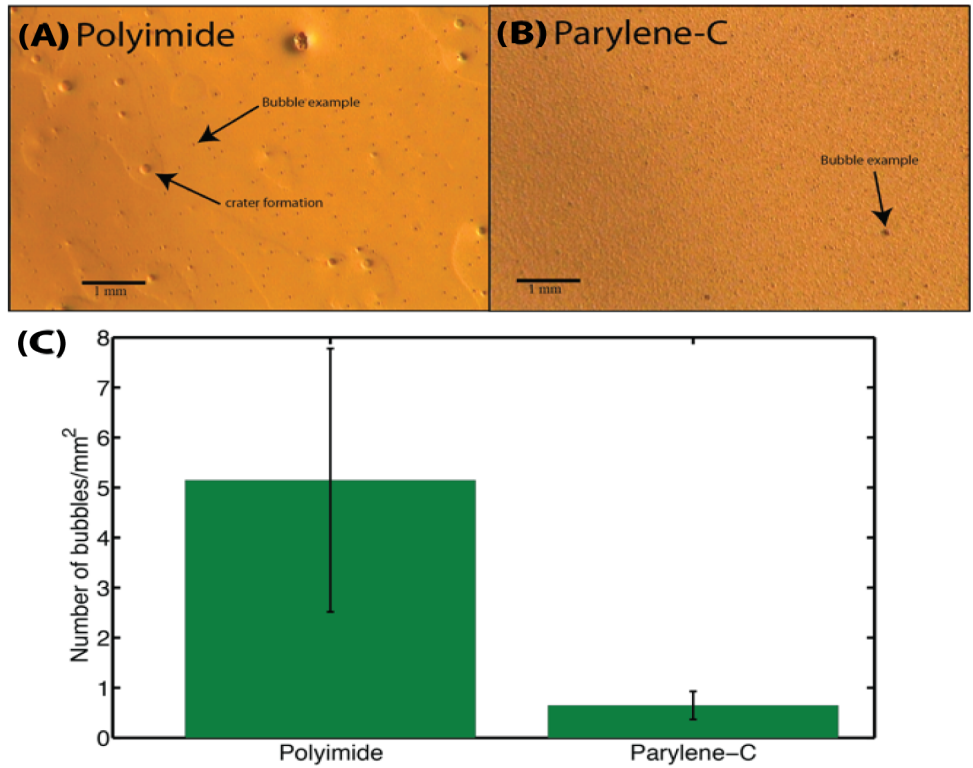

**Figure 5.** Representative image of a polyimide (A) and parylene-C (B) layer. The parylene-C layer had a slightly rough aluminium layer underneath and roughness is visible due to the conformal coating nature for parylene-C; bubble visualisation remains unaffected. (C) Bubble comparison test for a polyimide and parylene-C layer. Overall an 8:1 ratio is obtained for polyimide to parylene-C bubbles. For polyimide and parylene-C mean±sem values of 5.15±2.63 and 0.65± 0.28 bubbles/mm$^2$ are obtained respectively. The greater variation for polyimide further adds unpredictability about dielectric performance in respect to potential pitting.

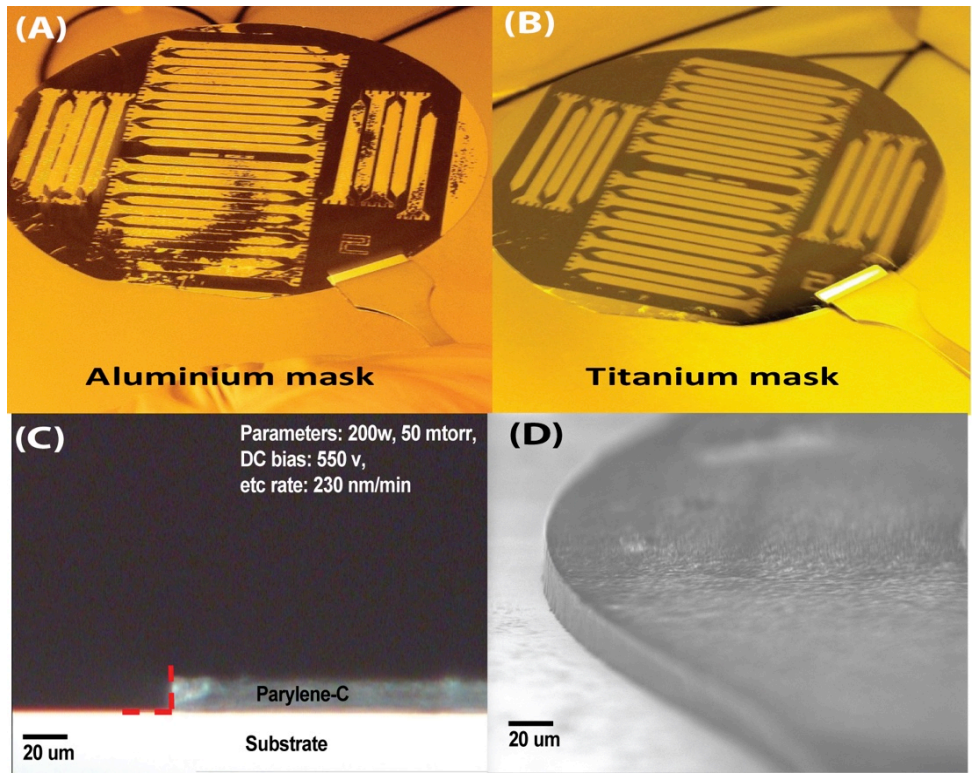

**Figure 6.** Optimisation of the thick layer Parylene-C etching. A Ti hard mask (B) was preferred over Al (A) due to poor adhesion. Highly vertical side walls were achieved as corroborated through optical (C) and scanning electron microscopy (SEM). SEM picture is of the rounded recording end for the sinusoidal probe (D).

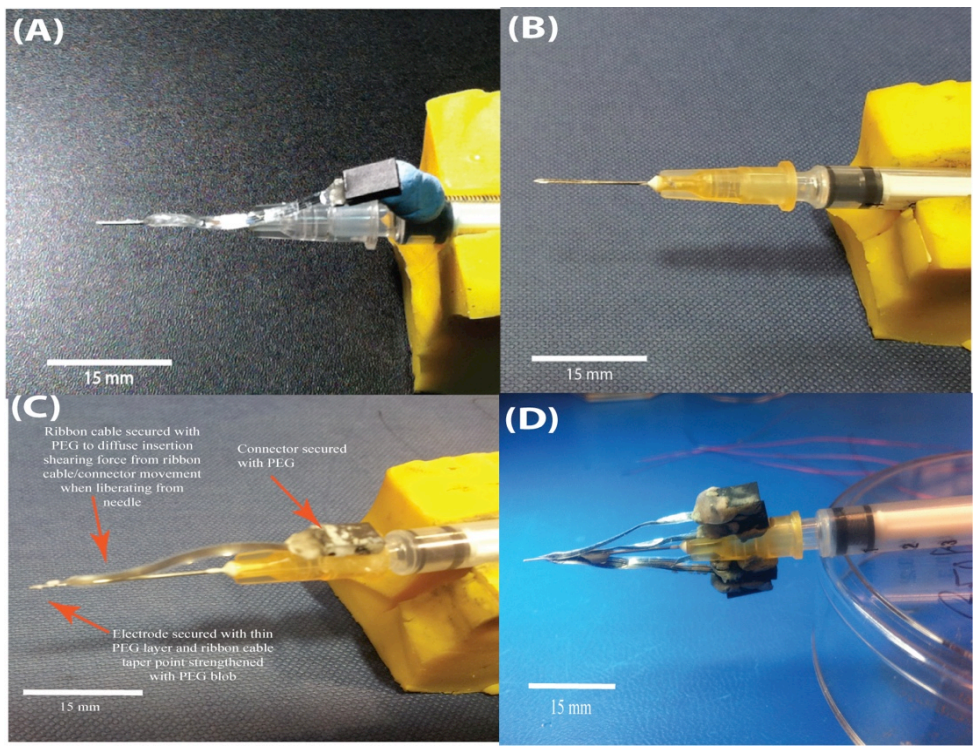

**Figure 7.** Evolution of a relatively cheap insertion carrier to aid insertion of the flexible sinusoidal probe into the brain. Initial 30G needles, epoxy filled were used (A) but the probe "slid" upon insertion. Machined syringes with a groove were then considered (B). The optimal device was a 25 G needle with a sharp steel electrode placed inside, with only the sharp steel electrode entering the brain (C). The connector and probe were attached temporarily with PEG. Further, we optimized the insertion device to carry up to 4 sinusoidal probes (D).

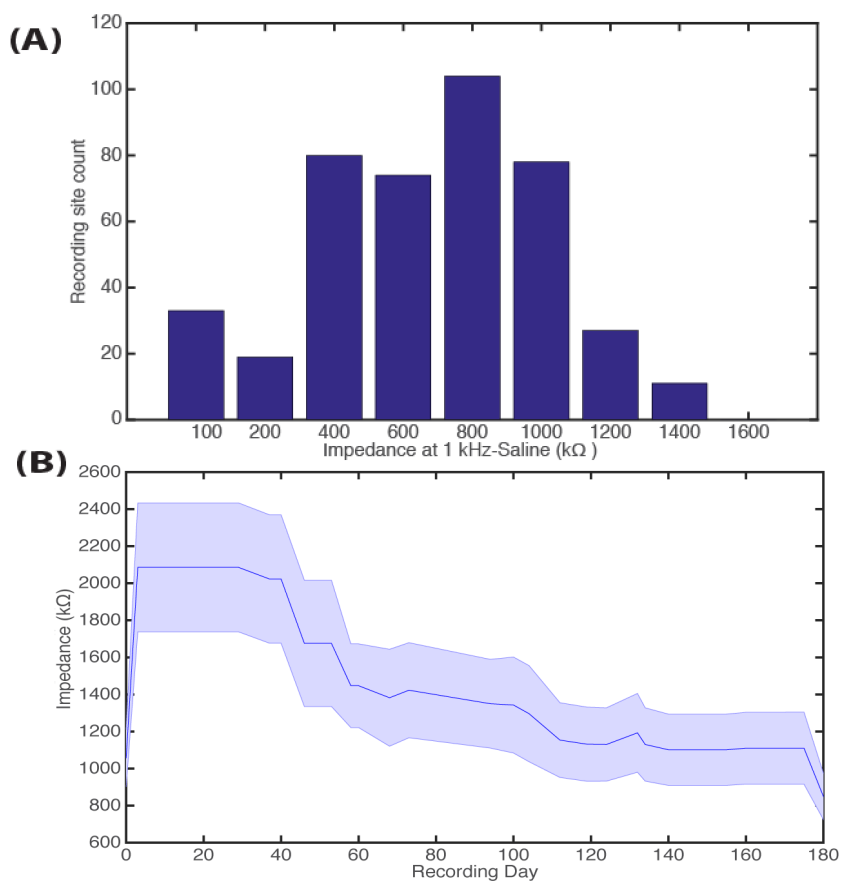

**Figure 8**. Histogram for device impedance of 426 recording sites, binned every 200 kΩ. We show that all impedance measurements of recording sites at 1 kHz in saline are at suitable impedances for single-unit recording (A). Electrode impedances over a 180 day indwelling period for 15 electrode recording sites the second generation probe. Day 0 and 180 equate to bench testing measuring impedances at 1 kHz in saline for pre-implant and, where possible, explanted devices, respectively. The days inbetween are a measure of the electrode impedance at the electrode-tissue interface in the rabbit cortex.

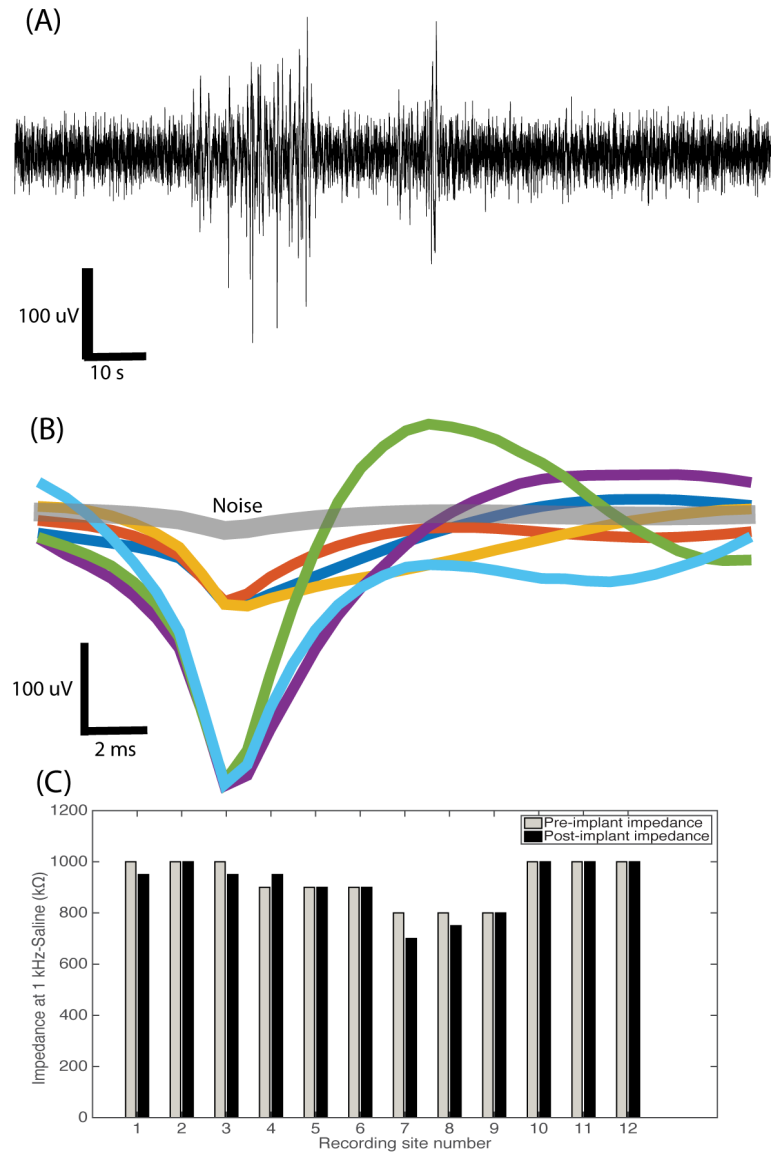

**Figure 9**. Sinusoidal probe acute neural recording example. (A) Example 250 Hz high passed raw signal acquired showing good neural activity. (B) Overlain average waveforms of isolated units (n=6) in comparison to noise waveforms (grey) isolated from a sinusoidal probe. we showed good: mean peak-to-peak amplitude (200+/-49 uV), Noise (33+/-7 uV) and SNR (6.0+/-0.5) (C) Impedance measurements pre- and post-implant for 12 recording sites (4 probes) showing no change in impedance (t-test: t=1, $P>0.05$).